\documentclass[aps,prl,reprint]{revtex4-1}
\usepackage{amsmath,graphicx}

\begin{document}
\title{Nanomechanical detection of the spin Hall effect}
\author{J. A. Boales}
\author{C. T. Boone}
\author{P. Mohanty}
\affiliation{Department of Physics, Boston University, 590 Commonwealth Avenue, Boston, MA 02215}

\begin{abstract}
The spin Hall effect creates a spin current in response to a charge current in a material that has strong spin-orbit coupling. The size of the spin Hall effect in many materials is disputed, requiring independent measurements of the effect. We develop a novel mechanical method to measure the size of the spin Hall effect, relying on the equivalence between spin and angular momentum. The spin current carries angular momentum, so the flow of angular momentum will result in a mechanical torque on the material. We determine the size and geometry of this torque and demonstrate that it can be measured using a nanomechanical device. Our results show that measurement of the spin Hall effect in this manner is possible and also opens possibilities for actuating nanomechanical systems with spin currents.
\end{abstract}

\maketitle

The spin Hall effect \cite{dyakonov1971current,hirsch1999spin}, which is the generation of a spin current in a material due to an applied charge current in the presence of strong spin-orbit coupling, has been proposed as a novel method of spin manipulation for spintronics applications. Spin currents generated by the spin Hall effect have been used to excite high- and low-frequency magnetic dynamics in nanostructures \cite{miron2011perpendicular, liu2012spin, liu2011review, demidov2011wide, demidov2012magnetic, haazen2013domain, duan2014spin, duan2014nanowire, emori2013current}, and may become useful for future low-power spintronic logic and storage devices \cite{jungwirth2012spin}. The spin Hall effect has been observed in a variety of materials with strong spin orbit coupling, including semiconductors such as Si and GaAs \cite{dash2009electrical, appelbaum2007electronic, kato2004observation, ehlert2012all}; graphene with adsorbed impurities \cite{balakrishnan2014giant}; heavy metals with strong spin-orbit coupling such as Pt, Ta and W \cite{valenzuela2006direct, kimura2007room, emori2013current, liu2012spin, pai2012spin}; and metals doped with large spin-orbit-coupled impurities \cite{niimi2011extrinsic, laczkowski2014experimental, niimi2012giant}. However, quantification of the SHE through fundamental parameters remains a challenge.

The figure-of-merit for SHE materials is the spin Hall angle, $\Theta_{SH}$, which is often stated as the proportionality between the magnitude of generated spin current and the magnitude of input charge current, $|J_s| = \left| \frac{\hbar}{2 e} \Theta_{SH} J_c \right|$ \cite{liu2011spin,khvalkovskiy2013matching}, where $J_s$ is the component of the spin current perpendicular to the charge current and $J_c$ is the charge current. Measurements and characterization of $\Theta_{SH}$ have, as yet, been limited to methods based on optical, electrical, and magnetic effects, which require knowledge of Kerr rotation coupling, metallic interfaces, magnetic properties, and spin diffusion parameters to quantify $\Theta_{SH}$ accurately \cite{kato2004observation,van2014optical}. As such, reported values of $\Theta_{SH}$ span orders of magnitude for materials such as Pt and Pd \cite{kimura2007room, niimi2013experimental, azevedo2011spin, castel2012platinum, zhang2013determination, tao2014spin, vlaminck2013dependence, weiler2014phase, du2014systematic, mosendz2010quantifying}. Open questions such as the dependence of $\Theta_{SH}$ on growth conditions, film thickness, impurity level, frequency, and other systematic parameters must be addressed both experimentally and theoretically. Since the inverse SHE is used to measure spin transport due to the spin Seebeck effect and spin pumping, accurate knowledge of the spin Hall angle is important for metrology. One intriguing new result implies that the spin Hall angle is complex-valued, resulting in a phase shift between an applied AC charge current and the resulting AC spin current \cite{weiler2014phase}. Additionally, the spin diffusion length in spin Hall materials may be correlated with the spin Hall angle, indicative of the same physics affecting both quantities. How these material parameters are related is now an area of intense study.

Because of the large discrepancies in measured spin Hall angle values using electrical techniques, independent methods of measuring the spin Hall effect are needed to accurately determine the spin Hall angle, the underlying mechanisms governing the effect, and the value of spin transport parameters such as the spin diffusion length. Though very promising for future study, as of yet, no quantitative results from metals have emerged from such a technique.

This paper describes a novel method of determining SHE properties by measuring the mechanical effects of a charge current applied to a SHE material using a nanomechanical resonator. This technique relies on the equivalence between spin angular momentum and mechanical angular momentum, and the fact that one is ultimately converted into the other. The equivalence between spin and mechanical angular momentum has been known since the initial experiments of Richardson \cite{richardson1908mechanical} and Einstein and de Haas \cite{einstein1915experimental}, and measurements of spin properties through mechanical effects have been performed to study spin injection from a ferromagnet into a nonferromagnetic metal \cite{mohanty2004spin, zolfagharkhani2008nanomechanical}. Recent discoveries of new spin-based effects, such as the spin Seebeck and spin Hall effects, in a variety of materials, from metals to semiconductors to insulators, have motivated us to study the associated spin physics micromechanically. We show that accurate determination of the spin Hall angle is possible with this method, which requires no \textit{a priori} knowledge of interfacial transparency, spin scattering mechanisms, and other unknown quantities. The results presented here also offer the possibility of novel excitation methods for nanomechanical devices using spin polarized currents generated by the SHE and others, including the spin Seebeck effect and topological insulators.

\begin{figure}
\includegraphics[width=\columnwidth]{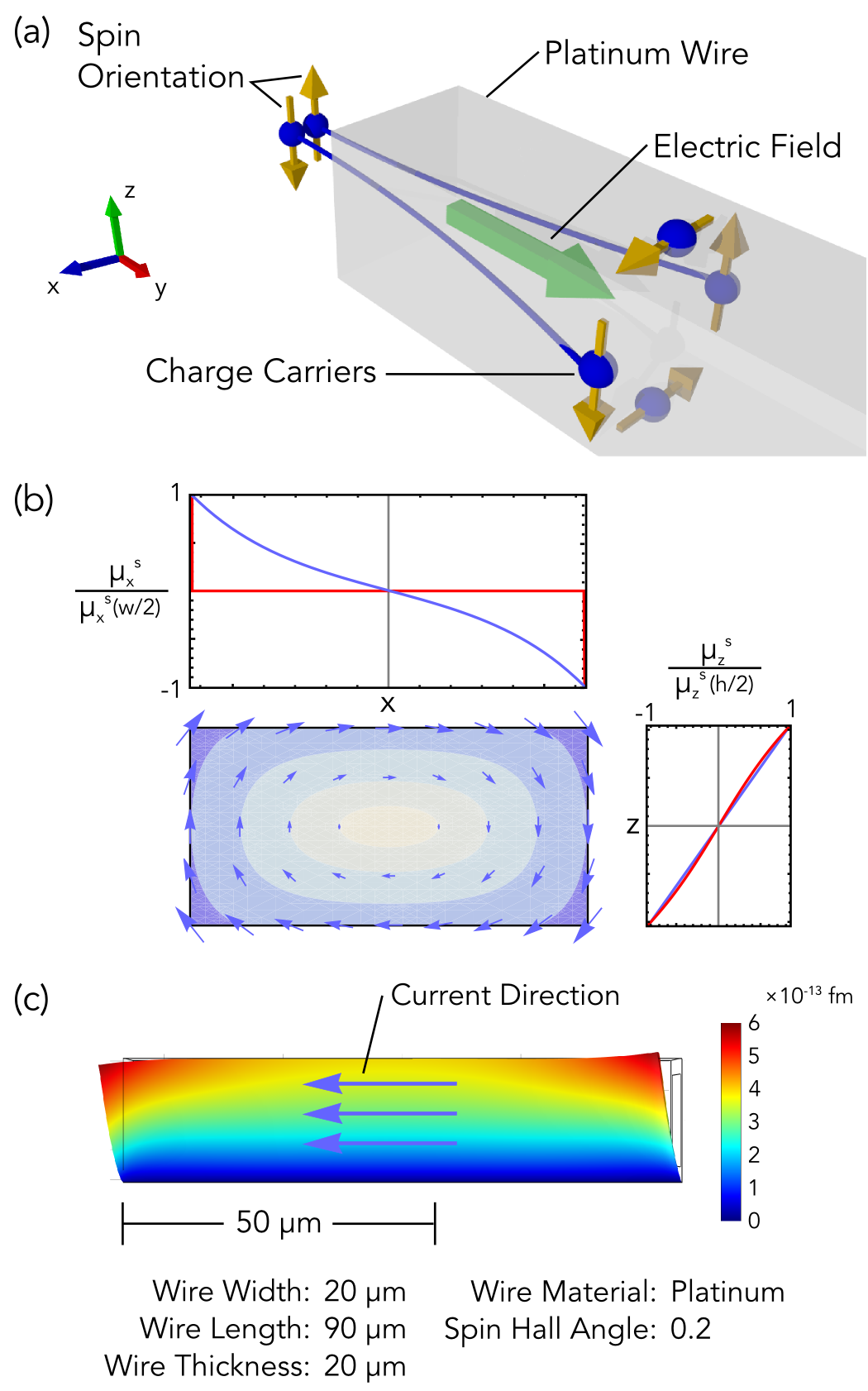}
\caption{\textbf{(a)} Schematic representation of the spin Hall effect. Oppositely polarized spins scatter in opposite directions, resulting in a spin accumulation at the wire boundary. \textbf{(b)} Graphical representation of the spin accumulation in a cross section of the wire. The blue lines and vector plot illustrate the accumulation for $\lambda_{sf}$ comparable to the wire size and the red lines represent a much smaller $\lambda_{sf}$. \textbf{(c)} Deflection (not to scale) caused in a platinum wire due to the spin-transport-induced torque density. The bottom of the wire is fixed in place while the rest may deform freely.}
\label{fig:fig1}
\end{figure}

The spin Hall effect is the generation of a pure spin current in response to a charge current in a material with substantial spin-orbit coupling. The spin current flow direction is perpendicular to the charge current and is polarized perpendicular to the directions of charge current flow and spin current flow. The effect stems from spin-orbit and impurity scattering which causes oppositely polarized spins to scatter in opposite directions, as illustrated in Figure \ref{fig:fig1}a.
This is described using the spin and charge drift-diffusion equations \cite{haney2013current, shchelushkin2005spin},
\begin{subequations}
\begin{gather}
Q_{ij} = -\frac{\hbar}{2 e^2} \sigma \nabla_i \mu_j^{(s)} - \frac{\hbar}{2 e^2} \Theta_{SH} \sigma \epsilon_{i j k} \nabla_k \mu^{(c)} \\
\nabla^2 \mathbf{\mu}^{(s)} = \frac{\mathbf{\mu}^{(s)}}{\lambda^2_{sf}} \\
J_c = -\sigma \vec{\nabla}\mu^{(c)} - \Theta_{SH} \sigma \vec{\nabla} \times \vec{\mu}^{(s)} \\
\nabla^2 \mu^{(c)} = 0
\end{gather}
\label{eq:spinDiff}
\end{subequations}

\vspace{-10.5pt}\noindent
where $\mu^{(c)}$ is the charge potential (defined such that $\frac{\mu^{(c)}}{e}$ is the applied voltage difference), $\mu_j^{(s)}$ is the spin accumulation (the difference in spin chemical potential between up and down spins) for spins aligned along direction $j$, $Q_{ij}$ is the spin current dyadic describing the flux along direction $i$ of spins aligned along direction $j$, $J_c$ is the charge current density, $\nabla_i$ is the derivative along coordinate direction $i$, $\epsilon_{ijk}$ is the antisymmetric Levi-Civita tensor, $\sigma$ is the electrical conductivity, and $\lambda_{sf}$ is the spin diffusion length. These equations describe coupling between spin currents and charge currents such that a charge (spin) potential becomes a source of spin (charge) current. Here, we are interested in the SHE, where we apply a charge current to generate a spin current. In our analysis later in the manuscript, we ignore inverse SHE corrections to the charge current which appear at second order in $\Theta_{SH}$. The steady-state approximation used to obtain Eqs. \ref{eq:spinDiff} is valid for time scales much longer than the spin relaxation time in metals (typically fs). Since the frequencies used in this analysis have time scales of order 1 $\mu$s, the steady-state approximation is appropriate. On these long time scales, other potential carriers of angular momentum (spin waves, etc.) have relaxed. Finally, we assume that hyperfine interactions are negligible.

The boundary conditions on the spin accumulation require that no spin current flows into the surrounding insulator media, i.e. the perpendicular component of Q equals zero at the structure's interfaces with insulating materials. This set of boundary conditions can be written in the form
\begin{equation}
\left.\nabla_i \mu_j^{(s)} \right|_\mathrm{boundary} = \left.-\Theta_{SH} \epsilon_{ijk} \nabla_k \mu^{(c)} \right|_\mathrm{boundary}.
\label{eq:BCs}
\end{equation}
We solve Eqs. \ref{eq:spinDiff} and \ref{eq:BCs} numerically using the COMSOL multiphysics package, for arbitrary geometries, to generate the steady-state spin accumulation and spin current density profiles for a given applied voltage. The net spin accumulation generated by the SHE is illustrated in Figure \ref{fig:fig1}b for a uniform charge current flowing in a rectangular thin film. The steady-state spin accumulation spirals around the center, increasing in magnitude toward the surfaces. Spins accumulate on all surfaces, polarized parallel to the surface and perpendicular to the charge current. Due to the lossy nature of spin transport in metals, the spin accumulation decays exponentially away from each boundary within the spin diffusion length, as shown in the insets of Figure \ref{fig:fig1}b. This figure illustrates the spin accumulation for spin diffusion lengths comparable to (blue line) and much smaller than (red line) the wire dimensions. Spin current, which is the spatial derivative of the spin accumulation, is therefore always flowing in the steady-state approximation and the absence of a sufficiently large magnetic field.

For a very long, rectangular geometry of thickness $h$, width $w$, and length $L$ carrying a current density $J_c$ (produced by a voltage $V_0$ over a length $L$), the diffusion equations are analytically solvable:
\begin{subequations}
\begin{gather}
\mu_x^{(s)} = \frac{\lambda_{sf} \Theta_{SH} J_c}{\sigma} \frac{\sinh\left(\frac{z}{\lambda_{sf}}\right)}{\cosh\left(\frac{h}{2 \lambda_{sf}}\right)} \\
\mu_y^{(s)} = 0 \\
\mu_z^{(s)} = -\frac{\lambda_{sf} \Theta_{SH} J_c}{\sigma}  \frac{\sinh\left(\frac{x}{\lambda_{sf}}\right)}{\cosh\left(\frac{w}{2 \lambda_{sf}}\right)} \\
Q_{xz} = \frac{\Theta_{SH} J_c \hbar}{2 e} \left( \frac{\cosh \left( \frac{x}{\lambda_{sf}} \right)}{\cosh\left( \frac{w}{2 \lambda_{sf}} \right)} - 1 \right) \\
Q_{zx} = -\frac{\Theta_{SH} J_c \hbar}{2 e} \left( \frac{\cosh \left( \frac{z}{\lambda_{sf}} \right)}{\cosh\left( \frac{h}{2 \lambda_{sf}} \right)} - 1 \right)
\end{gather}
\label{eq:analyticalResult}
\end{subequations}

\vspace{-5.5pt}\noindent
where the y-axis is defined along the length (parallel to the applied current), x is along the width, and z is along the thickness. All other components of Q are zero. This solution is consistent with prior results that have measured the interfacial spin accumulation as a function of thickness by its action on an adjacent ferromagnetic layer.
The flow of spins within the material is equivalent to a flow of angular momentum \cite{kovalev2007current}. Because total angular momentum is conserved, the change in angular momentum of the electrons must be compensated by a change in angular momentum of the lattice. This results in a mechanical torque on the material. The mechanical torque associated with the SHE can be calculated from conservation of angular momentum, assuming all spins gained and lost in the diffusive and SHE processes are converted to lattice angular momentum. For materials with no equilibrium magnetization, this assumption is valid. The local torque density, $\vec{\tau}$, is the local net change of angular momentum per unit volume. For SHE drive frequencies much smaller than the spin scattering rate ($<10^{10}$ $\mathrm{s^{-1}}$ \cite{mihajlovic2010surface, bass2007spin}), is  \cite{haney2010current}
\begin{equation}
\tau_j = \nabla_i Q_{ij}
\label{eq:torqueDensity}
\end{equation}
For a structure with dimensions much larger than the material's spin diffusion length, the net torque, $\vec{T} = \int\vec{\tau}\ dV$, is localized at the surfaces,
\begin{subequations}
\begin{gather}
T_{z,\mathrm{right/left}} = \pm \frac{\hbar}{2e} \Theta_{SH} J_c L h \\
T_{x,\mathrm{top/bottom}} = \pm \frac{\hbar}{2e} \Theta_{SH} J_c L w \label{eq:torqueTx}
\end{gather}
\label{eq:netTorque}
\end{subequations}

\vspace{-10.5pt}\noindent
where - (+) refers to the right (left) or top (bottom) surface. The torque averaged over the entire structure is zero, so that conservation of total angular momentum is maintained. However, the SHE creates a separation of angular momentum and thus introduces local mechanical torques. In this limit, the total torque at each surface is independent of the spin diffusion length, and contains only one free parameter: the spin Hall angle. The torque at the surface is aligned parallel to the surface and perpendicular to the applied current. This results in an effective force along each surface plane, compensated by an opposing force distribution in the interior.

The torque density can be converted to an effective force density profile (see Supplemental Material for derivation) which we apply to the SHE material to calculate the deformation.  For our numerical calculations, we use 10 nm for $\lambda_{sf}$ and 0.2 for $\Theta_{SH}$ \cite{bass2007spin, mosendz2010quantifying}. Other properties of Pt are summarized in the Supplemental material.

Figure \ref{fig:fig1}c shows the deformation of a Pt film, which is fixed at the bottom surface, induced by the SHE. The SHE causes the surfaces to move in the direction parallel to the current while the interior moves in the opposite direction. Any object mechanically coupled to the SHE film will feel the force applied at the contacting interface. 

We simulate the mechanical effect of the SHE on a microelectromechanical system (MEMS) to demonstrate its power as a detection method. Eqs. \ref{eq:spinDiff} and \ref{eq:BCs} are solved numerically using COMSOL with $\lambda_{sf}$, $\sigma$, $\Theta_{SH}$, and the applied voltage difference $V_0$ as input parameters, resulting in numerical profiles of the spin accumulation and spin current. The values of these parameters used in the simulations are summarized in the Supplemental material.

\begin{figure}
\includegraphics[width=\columnwidth]{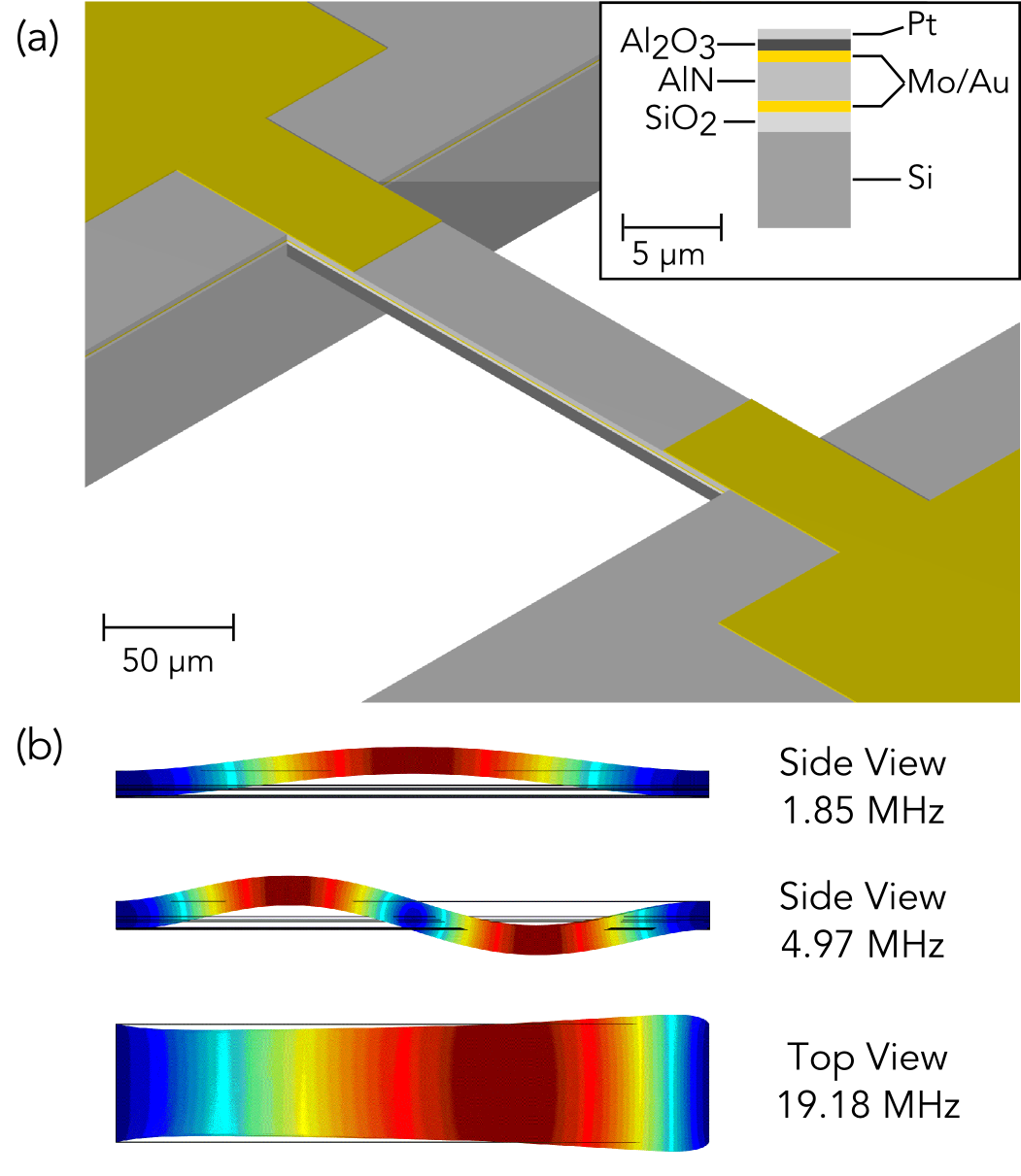}
\caption{\textbf{(a)} Representation of the doubly-clamped beam to be used in experiments (Pt and $\mathrm{Al_2O_3}$ layers not shown). The inset illustrates the layers used in constructing the beam. \textbf{(b)} Several of the lowest order mode shapes for the doubly-clamped beam according to a numerical simulation in COMSOL.}
\label{fig:fig2}
\end{figure}

The force profile, derived using Eq. \ref{eq:torqueDensity} and the process described in the Supplemental Material, is applied to the surface  of the MEMS structure, and the resulting deformation is calculated using linear elastic theory in COMSOL. Motivated by the force profile, which is approximately constant along the length of the SHE material, we choose an anchored beam, shown in Figure \ref{fig:fig2}a, as our MEMS for this demonstration.
In the future, other MEMS may be explored for device and detection optimization. The inset in Figure \ref{fig:fig2}a illustrates the beam's layered structure of Si(5,000) / $\mathrm{SiO_2}$(1,000) / Mo(600) / AlN(2,000) / Mo(600) / $\mathrm{Al_2O_3}$(60) / Pt(20) (thicknesses in nanometers). The bulk of the beam is Si / $\mathrm{SiO_2}$ for the mechanical properties and robustness, and the beam is capped with a Mo/AlN/Mo piezoelectric detection layer \cite{chen2011mechanical}. The piezoelectric layer allows for precise electrical measurements of the deformation of the MEMS, with the generated piezoelectric voltage being the measured output parameter. Other detection methods, including optical and magnetomotive, may in principle be employed instead. The thin $\mathrm{Al_2O_3}$ layer electrically separates the Pt from the Mo to avoid parasitic currents. The beam used in these simulations is 200 microns long and 40 microns wide. For dynamic simulations we include Rayleigh damping \cite{spears2009approach} at a value that produces a quality factor of 100 at 20 MHz. Figure \ref{fig:fig2}b shows representative mechanical eigenmodes of the beam, with eigenfrequencies in the rage of 10-200 MHz. A variety of eigenmodes with different symmetries exist, but only a limited number of modes will be preferentially excitable by the SHE, as we discuss later. This fact allows us to determine definitively that the mechanical excitation is due to the SHE.

\begin{figure}
\includegraphics[width=\columnwidth]{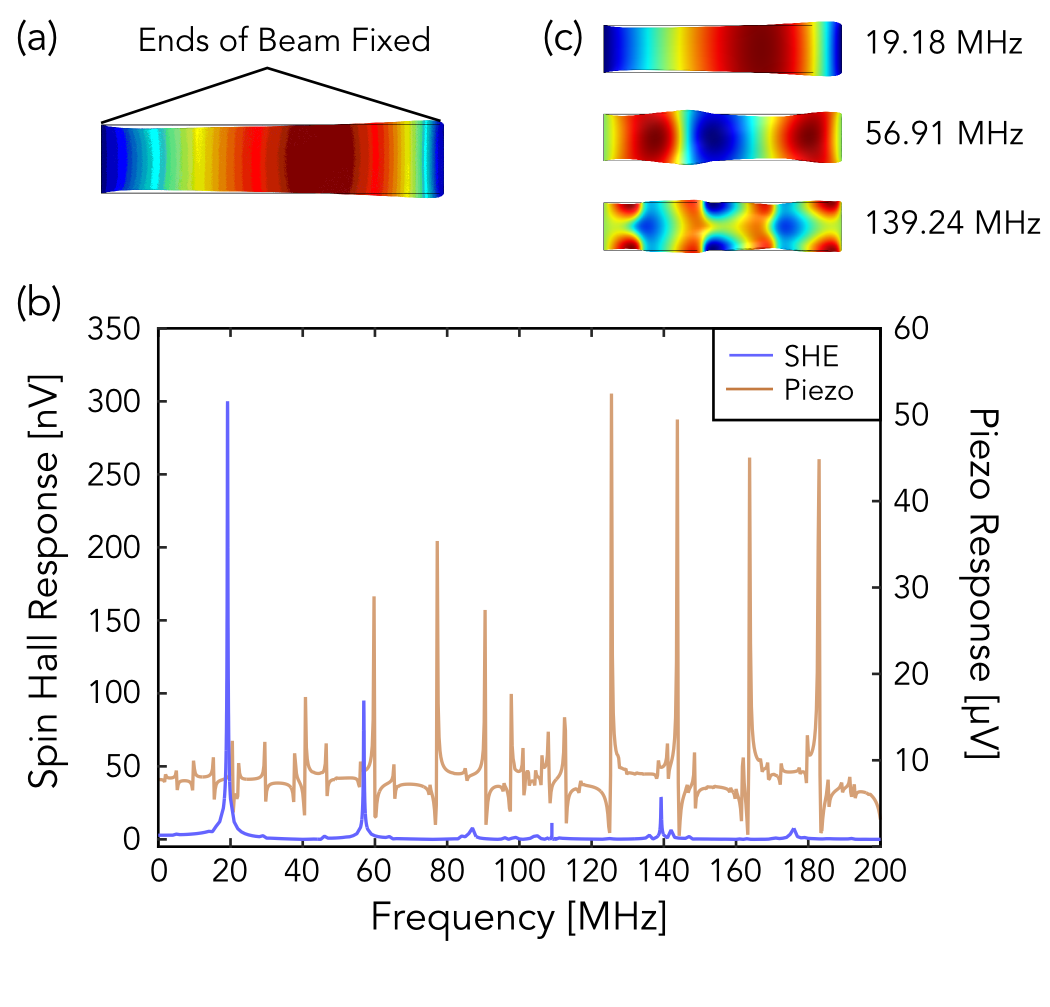}
\caption{\textbf{(a)} Deflection of the beam due to steady-state (dc) applied current. \textbf{(b)} Voltage response of oscillator when excited via the spin Hall effect (blue) and piezoelectrically (red) when driven by a 1 mV electric potential. \textbf{(c)} Mode shapes of the three eigenmodes most strongly excited by the spin Hall effect.}
\label{fig:fig3}
\end{figure}

We first analyze the response to an applied direct current (dc). The beam reaches a final equilibrium deformation shown in Figure \ref{fig:fig3}a.
In the linear elastic regime, the maximum deflection of the beam is approximately 0.28 fm/mA. The output voltage is in the range of nV for currents small enough to avoid Joule heating ($<$ 10 mA), and is strongly dependent on the thickness of the wire and the spin flip length. Unfortunately, the thermal noise is expected to be at least an order of magnitude larger than the DC signal due to the SHE at room temperature.

We also analyze the response under alternating current (ac) excitation. We apply the force profile at a given frequency and calculate the response at the input frequency, resulting in an effective scattering parameter for the SHE-MEMS device. Here, the force profile will only couple to the limited number of eigenmodes with strong spatial overlap. Figure \ref{fig:fig3}b shows the response due to SHE excitation and due to only piezoelectric excitation. In both cases, the driving potential has a magnitude of 1 mV. The piezoelectric contacts strongly excite many modes while the spin Hall effect only strongly excites three modes, at 19.18 MHz, 56.91 MHz, and 139.24 MHz. The three frequencies that couple strongly to the SHE have profiles shown in Figure \ref{fig:fig3}c. The two strongest peaks correspond to the two lowest order longitudinal modes, which involve vibration along the length of the wire, while the highest frequency peak corresponds to a more complicated mode shape. For the mode at 19.18 MHz, we can expect a response of approximately $300 \pm 92.5$ nV at 300 K or $300 \pm 11.6$ nV at 4 K. These measurements correspond to values for $\Theta_{SH}$ of $0.2 \pm 0.062$ at 300 K and $0.2 \pm 0.0075$ at 4 K. Since this analysis was performed using a 1 mV input potential and the output signal increases linearly with input, the signal size can easily be increased by at least two orders of magnitude without increasing the noise simply by increasing the driving voltage.  The sensitivity analysis detailed in the Supplemental Material.

In conclusion, we have calculated the mechanical response of a spin Hall material to an applied current. We showed that the mechanical response can be coupled to a MEMS and, the resulting signal is large enough to be measured with standard equipment. The resonant behavior of the SHE is unique, distinguishable by the different modes it excites. This technique can be used to measure the SHE at different temperatures, for a variety of materials, and can be extended to study other spin properties such as the surface states of topological insulators and magnetically-doped 3D topological insulators \cite{chang2013experimental}, edge states of quantum spin Hall insulators \cite{knez2012andreev}, spin waves in antiferromagnets, and materials  displaying the quantum anomalous Hall effect. Our results also offer prospects for new actuation methods for nanomechanical devices based on spin transfer effects. In-depth study of the spin Hall effect will reveal insights about spin-orbit coupling and the interplay between charge and spin transport in metals.

\bibliographystyle{apsrev4-1}
\bibliography{ConceptPaperFinal}

\newpage

\end{document}